\documentclass{article}
\makeatletter
\AtBeginDocument{\let\hl\@firstofone}
\makeatother
\usepackage{graphicx}% Include figure files
\usepackage{dcolumn}% Align table columns on decimal point
\usepackage{bm,slashed}% bold math
\usepackage{amssymb,amsmath}

\usepackage{url}
\DeclareMathOperator{\sgn}{sgn}
\makeatletter
\g@addto@macro{\UrlBreaks}{\UrlOrds}
\makeatother
\usepackage{float}
\usepackage{soul}
\usepackage{grffile}
\usepackage{mathtools}
\usepackage{stackengine}
\usepackage{vector}
\usepackage{upgreek}
\usepackage[OMLmathrm,OMLmathbf]{isomath}
\begin{document}
\title{Some Classical Models of Particles and Quantum Gauge~Theories}
\author{Andrey Akhmeteli\\LTASolid Inc., Houston, TX, USA\\akhmeteli@ltasolid.com}
\maketitle
\begin{abstract}
The article contains a review and new results of some mathematical models relevant to the interpretation of quantum mechanics and emulating well-known quantum gauge theories, such as scalar electrodynamics (Klein--Gordon--Maxwell electrodynamics), spinor electrodynamics (Dirac--Maxwell electrodynamics), etc. In these models, evolution is typically described by modified Maxwell equations.
In the case of scalar electrodynamics, the scalar complex wave function can be made real by a gauge transformation, the wave function can be algebraically eliminated from the equations of scalar electrodynamics, and the resulting modified Maxwell equations describe the independent evolution of the electromagnetic field. Similar results were obtained for spinor electrodynamics.
Three out of four components of the Dirac spinor can be algebraically eliminated from the Dirac equation, and the remaining component can be made real by a gauge transformation. A similar result was obtained for the Dirac equation in the Yang--Mills field. As quantum gauge theories play a central role in modern physics, the approach of this article may be sufficiently general.
One-particle wave functions can be modeled as plasma-like collections of a large number of particles and antiparticles. This seems to enable the simulation of quantum phase-space distribution functions, such as the Wigner distribution function, which are not necessarily non-negative.
\end{abstract}

\section{Introduction}

Recent progress in quantum information processing puts a new emphasis on foundations of quantum theory. However, there is currently no consensus on the interpretation of quantum theory (see, e.g.,~\cite{Vaidman}, where the lack of consensus is discussed from the point of view of the many-worlds interpretation), so this article may be of interest as it contains a review and new results of some relevant mathematical models emulating well-known quantum theories, such as scalar electrodynamics (Klein--Gordon--Maxwell electrodynamics), spinor electrodynamics (Dirac--Maxwell electrodynamics), etc. In~these models, evolution is typically described by modified Maxwell equations, so the models arguably do not require interpretation any more than classical electrodynamics and are classical in this~sense.

For example, in~the case of scalar electrodynamics, Schrödinger~\cite{Schroed} noticed in 1952 that the scalar complex wave function can be made real by a gauge transformation. It turned out~\cite{Akhm10,Akhmeteli-IJQI,Akhmeteli-EPJC} that, in~analogy with Dirac's ``new electrodynamics'' ~\cite{Dirac}, the~wave function can then be algebraically eliminated from the equations of scalar electrodynamics, and~the resulting modified Maxwell equations describe independent evolution of the electromagnetic field.
Similar results were obtained for spinor electrodynamics (which is more realistic) at the expense of the introduction of a complex four-potential of an electromagnetic field~\cite{Akhmeteli-EPJC}. There are reasons to believe that this limitation can be removed~\cite{Akhmetelilagr}. One of the reasons is that Schrödinger's observation was extended to the Dirac equation in electromagnetic fields: three out of four components of the Dirac spinor can be algebraically eliminated from the Dirac equation, and~the remaining component can be made real by a gauge transformation (\cite{Akhmeteli-JMP,Akhm2015,Akhmspr}; see also~in~\cite{Bagrov2014}, pp. 24--25,~\cite{Bagro}). Furthermore, a~similar result was obtained for the Dirac equation in the Yang--Mills field~\cite{Akhmeteliym}. As~quantum gauge theories play a central role in modern physics, the~approach of this article may be sufficiently general.
The resulting non-second-quantized theories describing the independent evolution of a gauge field can be embedded into quantum field theories using a generalization of the Carleman linearization~\cite{Kowalski,Kowalski2,nightnew}. For~a system of nonlinear partial differential equations, this procedure generates a system of linear equations in the Fock space, which looks like a second-quantized theory and is equivalent to the original nonlinear system on the set of solutions of the~latter.

Some people may find  the lack of explicit discreteness in the ``field-only'' models described above unappealing. Therefore, let us note that one real function is enough to describe matter in some well-established theories, such as the Dirac equation in an arbitrary electromagnetic field, which suggests some ``symmetry'' between positive and negative frequencies and, therefore, particles and antiparticles. As~a result, one-particle wave functions can be modeled as plasma-like collections of a large number of particles and antiparticles~\cite{Akhm10,Akhmeteli-EPJC,Akhm-Ent}. Another motivation for such models is the similarity of dispersion relations for quantum theories of matter, such as the Klein--Gordon equation, and~some simple plasma~models.

A criterion was developed for the approximation of continuous charge density distributions with integer total charge by discrete ones with integer charges based on the equality of partial Fourier sums. An~example of such approximation is computed using the homotopy continuation method. It was also proven that for any number of Fourier coefficients, one can always find a discrete distribution satisfying the criterion. An~example mathematical model is proposed. A~modification of the description for composite particles, such as nucleons or large molecules, describes them as collections including a composite particle and a large number of pairs of elementary particles and~antiparticles.

While it is not clear if there is some reality behind such a description, it can become a basis of an interesting model of quantum mechanics. For~example, it can offer an intuitive picture of the double-slit experiment. It also seems to enable the simulation of quantum phase-space distribution functions, such as the Wigner distribution function, which are not necessarily non-negative, whereas, according to Feynman~\cite{Feynm1982}, ``The only difference between a probabilistic classical world and the equations of the quantum world is that somehow or other it appears as if the probabilities would have to go negative, and~that we do not know, as~far as I know, how to~simulate''.

The plasma-like description is intuitive, uses some notions of quantum field theory, such as vacuum polarization, and~does not seem to have problems with wave function spreading, although~it implies that the wave function has something to do with charge distribution. There are some similarities with de Broglie--Bohm interpretation (Bohmian mechanics), especially for composite particles. Bohmian mechanics is sometimes considered not just as an interpretation but~also as another picture of quantum mechanics and a basis for computational methods~\cite{Sanz}. This can also be  a way to assess the plasma-like description of this~work.
\section{Methods and~Results}\label{s2}
\unskip

\subsection{Scalar~Electrodynamics}\label{n21}

This work heavily uses the results of~\cite{Dirac,Schroed}, so let us summarize some of them here. In~\cite{Dirac} (see also~\cite{Kamen} and references therein), Dirac considers the following conditions of stationary action for the free electromagnetic field Lagrangian subject to the constraint $A_\mu A^\mu=k^2$:
\begin{equation}\label{eq:pr1qr21}
\Box A_\mu-A^\nu_{,\nu\mu}=\lambda A_\mu,
\end{equation}
where $A^\mu$ is the potential of the electromagnetic field, and~$\lambda$ is a Lagrange multiplier. The~constraint represents a nonlinear gauge condition. One can assume that the conserved current in the right-hand side of Equation~(\ref{eq:pr1qr21}) is created by particles of mass $m$, charge $e$, and~momentum (not generalized momentum!) $p^\mu=\zeta A^\mu$, where $\zeta$ is a constant. If~these particles move in accordance with the Lorentz equations
\vspace{6pt}
\begin{equation}\label{eq:pr2qr21}
\frac{dp^\mu}{d\tau}=\frac{e}{m}F^{\mu\nu}p_\nu,
\end{equation}
where $F^{\mu\nu}=A^{\nu,\mu}-A^{\mu,\nu}$ is the electromagnetic field, and~$\tau$ is the proper time of the particle ($(d\tau)^2=dx^\mu dx_\mu$), then
\begin{equation}\label{eq:pr3qr21}
\frac{dp^\mu}{d\tau}=p^{\mu,\nu}\frac{dx_\nu}{d\tau}=\frac{1}{m}p_\nu p^{\mu,\nu}=\frac{\zeta^2}{m}A_\nu A^{\mu,\nu}.
\end{equation}
\hl{Due} to the constraint, $A_\nu A^{\nu,\mu}=0$, so
\begin{equation}\label{eq:pr4qr21}
A_\nu A^{\mu,\nu}=-A_\nu F^{\mu\nu}=-\frac{1}{\zeta}F^{\mu\nu}p_\nu.
\end{equation}
\hl{Therefore}, Equations~(\ref{eq:pr2qr21})--(\ref{eq:pr4qr21}) are consistent if $\zeta=-e$, and~then $p_\mu p^\mu=m^2$ implies $k^2=\frac{m^2}{e^2}$ (so far the discussion is limited to the case $-e A^0=p^0>0$).

Thus, Equation~(\ref{eq:pr1qr21}) with the gauge condition
\begin{equation}\label{eq:pr5qr21}
A_\mu A^\mu=\frac{m^2}{e^2}
\end{equation}
describes both the independent dynamics of the electromagnetic field and the consistent motion of charged particles in accordance with the Lorentz equations. The~words ``independent dynamics''  mean the following: if values of the spatial components $A^i$ of the potential ($i=1,2,3$) and their first derivatives with respect to $x^0$, $\dot{A}^i$, are known in the entire space at some moment in time ($x^0=const$), then $A^0$, $\dot{A}^0$ may be eliminated using Equation~(\ref{eq:pr5qr21}), $\lambda$ may be eliminated using Equation~(\ref{eq:pr1qr21}) for $\mu=0$ (the equation does not contain second derivatives with respect to $x^0$ for $\mu=0$), and~the second derivatives with respect to $x^0$, $\ddot{A}^i$, may be determined from Equation~(\ref{eq:pr1qr21}) for $\mu=1,2,3$.

In his comment on  Dirac's work, Schr\"{o}dinger~\cite{Schroed} considered the interacting  scalar charged field $\psi$ and electromagnetic field $F^{\mu\nu}$ with the Klein--Gordon--Maxwell equations of motion (scalar electrodynamics):
\begin{equation}\label{eq:pr7qr21}
(\partial^\mu+ieA^\mu)(\partial_\mu+ieA_\mu)\psi+m^2\psi=0,
\end{equation}
\begin{equation}\label{eq:pr8qr21}
\Box A_\mu-A^\nu_{,\nu\mu}=j_\mu,
\end{equation}
\begin{equation}\label{eq:pr9qr21}
j_\mu=ie(\psi^*\psi_{,\mu}-\psi^*_{,\mu}\psi)-2e^2 A_\mu\psi^*\psi+j_\mu^{ext},
\end{equation}
where $j_\mu^{ext}$ is some conserved external current (added by the present author for generality).
The metric signature is $(+,-,-,-)$.

The complex charged matter field $\psi$ in scalar electrodynamics (Equations (\ref{eq:pr7qr21})--(\ref{eq:pr9qr21})) can be made real by a gauge transform (at least locally), and~the equations of motion in the relevant gauge (unitary gauge) for the transformed four-potential of electromagnetic field $B^{\mu}$ and real matter field $\varphi$ are as follows~\cite{Schroed}:
\begin{equation}\label{eq:pr10qr21}
\Box\varphi-(e^2 B^\mu B_\mu-m^2)\varphi=0,
\end{equation}
\begin{equation}\label{eq:pr11qr21}
\Box B_\mu-B^\nu_{,\nu\mu}=j_\mu,
\end{equation}
\begin{equation}\label{eq:pr12qr21}
j_\mu=-2e^2 B_\mu\varphi^2+j_\mu^{ext}.
\end{equation}

Schr\"{o}dinger emphasized two circumstances. Firstly, except~for the missing constraint, the~equations for the electromagnetic potentials coincide with Equation~(\ref{eq:pr1qr21}) (if we replace $B_\mu$ with $A_\mu$ and $-2e^2\varphi^2$ with $\lambda$). Secondly, the~fact that the scalar field can be made real by a change in gauge, although~easy to understand, contradicts the widespread belief about charged fields requiring complex representation. This consideration may be regarded as a part of the rationale for the present work, where real fields are used to describe charged~particles.

The following unexpected result was proven in~\cite{Akhmeteli-IJQI}: the equations obtained from Equations~(\ref{eq:pr10qr21})--(\ref{eq:pr12qr21}) after natural elimination of the matter field form a closed system of partial differential equations and thus describe independent dynamics of the electromagnetic field. The~detailed wording is as follows: if components of the four-potential of the electromagnetic field and their first derivatives with respect to time are known in the entire space at some time point, the~values of their second derivatives with respect to time can be calculated for the same time point, so the Cauchy problem can be posed, and~integration yields the four-potential in the entire~space-time.

To eliminate the matter field $\varphi$ from Equations~(\ref{eq:pr10qr21})--(\ref{eq:pr12qr21}), let us use a substitution $\Phi=\varphi^2$ first. For~example, as~\begin{equation}\label{eq:pr1qqr21}
\Phi_{,\mu}=2\varphi\varphi_{,\mu},
\end{equation}
we obtain
\begin{equation}\label{eq:pr2qqr21}
\Phi_{,\mu}^{,\mu}=2\varphi^{,\mu}\varphi_{,\mu}+2\varphi\varphi^{,\mu}_{,\mu}=
\frac{1}{2}\frac{\Phi^{,\mu}\Phi_{,\mu}}{\Phi}+2\varphi\varphi^{,\mu}_{,\mu}.
\end{equation}
\hl{Multiplying} Equation~(\ref{eq:pr10qr21}) by $2\varphi$, we obtain the following equations in terms of $\Phi$ instead of Equations~(\ref{eq:pr10qr21})--(\ref{eq:pr12qr21}):
\begin{equation}\label{eq:pr3qqr21}
\Box\Phi-\frac{1}{2}\frac{\Phi^{,\mu}\Phi_{,\mu}}{\Phi}-2(e^2 B^\mu B_\mu-m^2)\Phi=0,
\end{equation}
\begin{equation}\label{eq:pr4qqr21}
\Box B_\mu-B^\nu_{,\nu\mu}=-2e^2 B_\mu\Phi+j_\mu^{ext}.
\end{equation}
\hl{To} prove that these equations describe the independent evolution of the electromagnetic field $B^\mu$, it is sufficient to prove that if components $B^\mu$ of the potential and their first derivatives with respect to $x^0$ ($\dot{B}^\mu$) are known in the entire space at some time point $x^0=\rm{const}$  (that means that all spatial derivatives of these values are also known in the entire space at that time point), Equations~(\ref{eq:pr3qqr21}) and (\ref{eq:pr4qqr21}) yield the values of their second derivatives, $\ddot{B}^\mu$, for~the same value of $x^0$. Indeed, $\Phi$ can be eliminated using Equation~(\ref{eq:pr4qqr21}) for $\mu=0$, as~this equation does not contain $\ddot{B}^\mu$ for this value of $\mu$:
\begin{eqnarray}\label{eq:pr5qqr21}
\Phi=(-2e^2 B_0)^{-1}(\Box B_0-B^\nu_{,\nu 0}-j_0^{ext})=
(-2e^2 B_0)^{-1}(B^{,i}_{0,i}-B^i_{,i 0}-j_0^{ext})
\end{eqnarray}
(\hl{Greek} indices in the Einstein sum convention run from $0$ to $3$, and~Latin indices run from $1$ to $3$).
Then $\ddot{B}^i$ ($i=1,2,3$) can be determined by the substitution of Equation~(\ref{eq:pr5qqr21}) into Equation~(\ref{eq:pr4qqr21}) for $\mu=1,2,3$:
\begin{equation}\label{eq:pr6qqr21}
\ddot{B}_i=-B^{,j}_{i,j}+B^\nu_{,\nu i}+(B_0)^{-1} B_i(B^{,j}_{0,j}-B^j_{,j 0}-j_0^{ext})+j_i^{ext}.
\end{equation}
\hl{Thus}, to~complete the proof, we only need to find $\ddot{B}^0$. Conservation of current implies
\begin{equation}\label{eq:pr7qqr21}
0=(B^\mu \Phi)_{,\mu}=B^\mu_{,\mu}\Phi+B^\mu\Phi_{,\mu}.
\end{equation}
\hl{This} equation determines $\dot{\Phi}$, as~spatial derivatives of $\Phi$ can be found from Equation~(\ref{eq:pr5qqr21}). Differentiation of this equation with respect to $x^0$ yields
\begin{eqnarray}\label{eq:pr8qqr21}
\nonumber
0=(\ddot{B}^0+\dot{B}^i_{,i})\Phi+(\dot{B}^0+B^i_{,i})\dot{\Phi}+\\
\dot{B}^0\dot{\Phi}+\dot{B}^i\Phi_{,i}+B^0\ddot{\Phi}+B^i\dot{\Phi}_{,i}.
\end{eqnarray}
\hl{After} substitution of $\Phi$ from Equation~(\ref{eq:pr5qqr21}), $\dot{\Phi}$ from Equation~(\ref{eq:pr7qqr21}), and~$\ddot{\Phi}$ from  Equation~(\ref{eq:pr3qqr21}) into  Equation~(\ref{eq:pr8qqr21}), the~latter equation determines $\ddot{B^0}$ as a function of $B^\mu$, $\dot{B}^\mu$ and their spatial derivatives (again, spatial derivatives of $\Phi$ and $\dot{\Phi}$ can be found from the expressions for $\Phi$ and $\dot{\Phi}$ as functions of $B^\mu$ and $\dot{B}^\mu$). Thus, if~$B^\mu$ and $\dot{B}^\mu$ are known in the entire space at a certain value of $x^0$, then $\ddot{B}^\mu$ can be calculated for the same $x^0$, so integration yields $B^\mu$ in the entire space-time. Therefore, we  have the independent dynamics of the electromagnetic~field.

\subsection{Dirac Equation as an Equation for One~Function}\label{n22}

It turns out, however, that Schr\"{o}dinger's results also hold in the case of spinor electrodynamics. In~general, four complex components of the Dirac spinor function cannot be made real by a single gauge transform, but~three complex components out of four can be eliminated from the Dirac equation in a general case, yielding a fourth-order partial differential equation for the remaining component, and the latter can be made real by a gauge transform~\cite{Akhmeteli-JMP}. The~resulting two equations for one real component can be replaced by one equation plus the current conservation equation, and~the latter can be derived from the Maxwell~equations.

Spinor electrodynamics is a more realistic theory than scalar electrodynamics, so it seems important that the charged field of spinor electrodynamics can also be described by one real~function.

Let us consider the Dirac equation:
\begin{equation}\label{eq:pr25qr22}
(i\slashed{\partial}-\slashed{A})\psi=\psi,
\end{equation}
where, e.g.,~$\slashed{A}=A_\mu\gamma^\mu$ (the Feynman slash notation). For~the sake of simplicity, a~system of units is used where $\hbar=c=m=1$, and~the electric charge $e$ is included in $A_\mu$ ($eA_\mu \rightarrow A_\mu$).
In the chiral representation of $\gamma$-matrices~\cite{Itzykson}
\begin{equation}\label{eq:d1qr22}
\gamma^0=\left( \begin{array}{cc}
0 & -I\\
-I & 0 \end{array} \right),\gamma^i=\left( \begin{array}{cc}
0 & \sigma^i \\
-\sigma^i & 0 \end{array} \right),
\end{equation}
where index $i$ runs from 1 to 3, and~$\sigma^i$ are the Pauli matrices.
If $\psi$ has components
\begin{equation}\label{eq:d2qr22}
\psi=\left( \begin{array}{c}
\psi_1\\
\psi_2\\
\psi_3\\
\psi_4\end{array}\right),
\end{equation}
the Dirac Equation~(\ref{eq:pr25qr22}) can be written in components as follows:
\begin{eqnarray}\label{eq:d3qr22}
(A^0+A^3)\psi_3+(A^1-i A^2)\psi_4+
i(\psi_{3,3}-i\psi_{4,2}+\psi_{4,1}-\psi_{3,0})=\psi_1,
\end{eqnarray}
\begin{eqnarray}\label{eq:d4qr22}
(A^1+i A^2)\psi_3+(A^0-A^3)\psi_4-
i(\psi_{4,3}-i\psi_{3,2}-\psi_{3,1}+\psi_{4,0})=\psi_2,
\end{eqnarray}
\begin{eqnarray}\label{eq:d5qr22}
(A^0-A^3)\psi_1-(A^1-i A^2)\psi_2-
i(\psi_{1,3}-i\psi_{2,2}+\psi_{2,1}+\psi_{1,0})=\psi_3,
\end{eqnarray}
\begin{eqnarray}\label{eq:d6qr22}
-(A^1+i A^2)\psi_1+(A^0+A^3)\psi_2+
i\psi_{2,3}+\psi_{1,2}-i(\psi_{1,1}+\psi_{2,0})=\psi_4.
\end{eqnarray}
\hl{Obviously}, Equations~(\ref{eq:d5qr22}) and (\ref{eq:d6qr22}) can be used to express components $\psi_3,\psi_4$ via $\psi_1,\psi_2$ and eliminate them from Equations~(\ref{eq:d3qr22}) and (\ref{eq:d4qr22}) (cf.~\cite{Feygel,Laporte}). The~resulting equations for $\psi_1$ and $\psi_2$ are as follows:
\begin{eqnarray}\label{eq:d7qr22}
\nonumber
-\psi_{1,\mu}^{,\mu}+\psi_2(-i A^1_{,3}-A^2_{,3}+A^0_{,2}+A^3_{,2}+
i(A^0_{,1}+A^3_{,1}+A^1_{,0})+A^2_{,0})+\\
+\psi_1(-1+A^{\mu} A_{\mu}-i A^{\mu}_{,\mu}+i A^0_{,3}-A^1_{,2}+A^2_{,1}+i A^3_{,0})-2 i A^{\mu}\psi_{1,\mu}=0,
\end{eqnarray}
\begin{eqnarray}\label{eq:d8qr22}
\nonumber
-\psi_{2,\mu}^{,\mu}+i\psi_1( A^1_{,3}+i A^2_{,3}+i A^0_{,2}-i A^3_{,2}+
A^0_{,1}-A^3_{,1}+A^1_{,0}+i A^2_{,0})+\\
+\psi_2(-1+A^{\mu} A_{\mu}-i( A^{\mu}_{,\mu}+A^0_{,3}+i A^1_{,2}-i A^2_{,1}+ A^3_{,0}))-2 i A^{\mu}\psi_{2,\mu}=0.
\end{eqnarray}

Equations~(\ref{eq:d7qr22}) and (\ref{eq:d8qr22}) can be rewritten as follows:
\begin{eqnarray}\label{eq:d72qr22}
-\left(\Box'+i F^3\right)\psi_1-\left(i F^1+F^2\right)\psi_2=0,
\end{eqnarray}
\begin{eqnarray}\label{eq:d82qr22}
-\left(\Box'-i F^3\right)\psi_2-\left(i F^1-F^2\right)\psi_1=0,
\end{eqnarray}
where
\begin{equation}\label{eq:fi}
F^i=E^i+i H^i,
\end{equation}
electric field $E^i$ and magnetic field $H^i$ are defined by the standard formulae
\begin{eqnarray}\label{eq:d8n1qr22}
F^{\mu\nu}=A^{\nu,\mu}-A^{\mu,\nu}=\left( \begin{array}{cccc}
0 & -E^1 & -E^2 & -E^3\\
E^1 & 0 & -B^3 & B^2\\
E^2 & B^3 & 0 & -B^1\\
E^3 &-B^2 & B^1 & 0  \end{array} \right),
\end{eqnarray}
and the modified d'Alembertian $\Box'$ is defined as follows:
\begin{eqnarray}\label{eq:d8n2qr22}
\Box'=\partial^\mu\partial_\mu+2 i A^\mu\partial_\mu+i A^\mu_{,\mu}-A^\mu A_\mu+1.
\end{eqnarray}

As Equation~(\ref{eq:d72qr22}) contains $\psi_2$, but~not its derivatives, it can be used to express $\psi_2$ via $\psi_1$, eliminate it from Equation~(\ref{eq:d82qr22}) and obtain an equation of the fourth order for $\psi_1$:
\begin{eqnarray}\label{eq:deq}
\left(\left(\Box'-i F^3\right)\left(i F^1+F^2\right)^{-1}\left(\Box'+i F^3\right)-i F^1+F^2\right)\psi_1=0,
\end{eqnarray}

It should be noted that the coefficient at $\psi_2$ in Equation~(\ref{eq:d7qr22}) is gauge-invariant (it can be expressed via electromagnetic fields). While this elimination could not be performed for zero electromagnetic field, this does not look like a serious limitation, as~in reality there always exist electromagnetic fields in the presence of charged fields, although~they may be very small. It is not clear how free field being a special case is related to the divergences in quantum electrodynamics. It should also be noted that the above procedure could be applied to any component of the spinor function, not just to $\psi_1$. The~above results are presented in a more symmetric form in the next~subsection.

Using a gauge transform, it is possible to make $\psi_1$ real (at least locally). Then the real and the imaginary parts of Equation~(\ref{eq:d8qr22}) after the substitution of the expression for $\psi_2$ will present two equations for $\psi_1$. However, it is possible to construct just one equation for $\psi_1$ in such a way that the system containing this equation and the current conservation equation will be equivalent to Equation~(\ref{eq:d8qr22}).

\subsection{Algebraic Elimination of Components from the Dirac Equation in a General~Form}\label{n23}

To find a manifestly covariant form of Equation~(\ref{eq:deq}) \cite{Akhm2015,Akhmspr}, let us start again with the Dirac Equation~(\ref{eq:pr25qr22}). However, this time we do not choose a specific representation of the $\gamma$-matrices, but~only assume that the set of $\gamma$-matrices satisfies the standard hermiticity conditions~\cite{Itzykson}:
\begin{equation}\label{eq:li5aqr27}
\gamma^{\mu\dag}=\gamma^0\gamma^\mu\gamma^0, \gamma^{5\dag}=\gamma^5,
\end{equation}
where
\begin{equation}\label{eq:gamma5}
\gamma^5=i\gamma^0\gamma^1\gamma^2\gamma^3.
\end{equation}
\hl{Then} a charge conjugation  matrix $C$ can be chosen in such a way~\cite{Bogol,Schweber} that
\begin{equation}\label{eq:li5qr27}
C\gamma^\mu C^{-1}=-\gamma^{\mu T}, C\gamma^5 C^{-1}=\gamma^{5 T}, C\sigma^{\mu\nu}C^{-1}=-\sigma^{\mu\nu T},
\end{equation}
\begin{equation}\label{eq:li6qr27}
C^T=C^\dag=-C, CC^\dag=C^\dag C=I, C^2=-I,
\end{equation}
where the superscript $T$ denotes transposition, and~$I$ is the unit matrix.
Multiplying both sides of Equation~(\ref{eq:pr25qr22}) by $(i\slashed{\partial}-\slashed{A})$ from the left and using Equation~(\ref{eq:d8n1qr22}) and notation
\begin{equation}\label{eq:li2qr27}
\sigma^{\mu\nu}=\frac{i}{2}[\gamma^\mu,\gamma^\nu],
\end{equation}
we obtain (see details in~\cite{Akhm2015,Akhmspr}):
\begin{eqnarray}\label{eq:li1qr27}
\nonumber
\psi=(i\gamma^\nu\partial_\nu-A_\nu\gamma^\nu)(i\gamma^\mu\partial_\mu-A_\mu\gamma^\mu)\psi=
\\
(-\partial^\mu\partial_\mu-2 i A^\mu\partial_\mu-i A^\mu_{,\mu}+A^\mu A_\mu-\frac{1}{2}F_{\nu\mu}\sigma^{\nu\mu})\psi.
\end{eqnarray}
(\hl{A} similar equation can be found in the original article by Dirac~\cite{Dirac28}. Feynman and Gell-Mann~\cite{Feygel} used a similar equation to eliminate two out of four components of the Dirac spinor function. See also an earlier article~\cite{Laporte}). We obtain:
\begin{equation}\label{eq:li3qr27}
(\Box'+F)\psi=0,
\end{equation}
where the modified d'Alembertian $\Box'$ is defined by Equation~(\ref{eq:d8n2qr22})
and
\begin{equation}\label{eq:li4qr27}
F=\frac{1}{2}F_{\nu\mu}\sigma^{\nu\mu}.
\end{equation}
\hl{Note} that $\Box'$ and $F$ are manifestly relativistically~covariant.

Let us choose a component of the Dirac spinor $\psi$ in the form $\bar{\xi}\psi$, where $\xi$ is a constant spinor  (so it does not depend on the space-time coordinates $x=(x^0,x^1,x^2,x^3)$, and~$\partial_\mu\xi\equiv 0$), and~multiply  both sides of Equation~(\ref{eq:li3qr27}) by $\bar{\xi}$ from the left:
\begin{equation}\label{eq:li7qr27}
\Box'(\bar{\xi}\psi)+\bar{\xi}F\psi=0.
\end{equation}
\hl{To} derive an equation for only one component $\bar{\xi}\psi$, we need to express $\bar{\xi}F\psi$ via $\bar{\xi}\psi$. To~simplify this task, we demand that $\xi$ is an eigenvector of $\gamma^5$ (in other words, $\xi$ is either right-handed or left-handed). This condition is Lorentz-invariant~\cite{Akhm2015,Akhmspr}.

Eigenvalues of $\gamma^5$ equal either $+1$ or $-1$, so $\gamma^5\xi=\pm\xi$. The~linear subspace of eigenvectors of $\gamma^5$ with the same eigenvalue as $\xi$ is two-dimensional, so we can choose another constant spinor $\eta$ that is an eigenvector of $\gamma^5$  with the same eigenvalue as $\xi$ in such a way that $\xi$ and $\eta$ are linearly independent. This choice is Lorentz-covariant~\cite{Akhm2015,Akhmspr}.

Eventually, as~proven in~\cite{Akhm2015,Akhmspr}, we obtain:
\begin{equation}\label{eq:li25qr27}
(((\bar{\xi}\eta^c)\Box'-\bar{\xi}F\eta^c)(\bar{\xi}F\xi^c)^{-1}((\bar{\xi}\eta^c)\Box'
+\bar{\xi}F\eta^c)+\bar{\eta}F\eta^c)(\bar{\xi}\psi)=0.
\end{equation}
\hl{This} equation looks more complex than Equation~(\ref{eq:deq}), but~it is much more general and manifestly relativistically covariant. It can be slightly simplified if we require that $\xi$ and $\eta$ are normalized in such a way that $\bar{\xi}\eta^c=1$ (this condition also implies linear independence of $\xi$ and $\eta$). Further simplification can be achieved using the following notation for components of the electromagnetic field:
\begin{equation}\label{eq:simp}
f_{\alpha \beta}=\bar{\alpha}F\beta^c,
\end{equation}
where $\alpha$ and $\beta$ are some Dirac spinors. Then we obtain the following instead of \mbox{Equation~(\ref{eq:li25qr27}):}
\begin{equation}\label{eq:sim2}
((\Box'-f_{\xi\eta})f_{\xi\xi}^{-1}(\Box'
+f_{\xi\eta})+f_{\eta\eta})(\bar{\xi}\psi)=0.
\end{equation}

A different choice of $\eta$ yields an equivalent equation~\cite{Akhm2015,Akhmspr}.

This equation for one component $\bar{\xi}\psi$ is generally equivalent to the Dirac equation (if $\bar{\xi}F\xi^c\slashed{\equiv}0$): on the one hand, it was derived from the Dirac equation, and on~the other hand, the~Dirac spinor $\psi$ can be restored if its component $\bar{\xi}\psi$ is known~\cite{Akhm2015,Akhmspr}.

\subsection{An Approach to Elimination of the Spinor Field from the Equations of Spinor~Electrodynamics}\label{n24}

The following suggests that differential algebraic elimination of the spinor field from the equations of spinor electrodynamics is possible~\cite{Akhmetelilagr}. The~equations of spinor electrodynamics are as follows:
\begin{equation}\label{eq:pr25qr24}
(i\slashed{\partial}-\slashed{A}-1)\psi=0,
\end{equation}
\begin{equation}\label{eq:pr25aqr24}
\bar{\psi}(i\overleftarrow{\slashed{\partial}}+\slashed{A}+1)=0,
\end{equation}
\begin{equation}\label{eq:pr26qr24}
\Box A_\mu-A^\nu_{,\nu\mu}=e^2\bar{\psi}\gamma_\mu\psi
\end{equation}
(see notation at the beginning of Section \ref{n22}). The~chiral representation of $\gamma$-matrices (Equation (\ref{eq:d1qr22})) is~used.

Now, to~fix the gauge, let us require that $\psi_1$, the~first component of the spinor field
\begin{equation}\label{eq:d2}
\psi=\left( \begin{array}{c}
\psi_1\\
\psi_2\\
\psi_3\\
\psi_4\end{array}\right),\bar{\psi}=\psi^\dagger\gamma_0= (-\psi_3^*,-\psi_4^*,-\psi_1^*,-\psi_2^*),
\end{equation} is real: $\psi_1=\psi_1^*$. As~$\psi_1$ generally does not vanish identically, this does fix the gauge (in a general case).

Using the elimination process of  Section \ref{n22}, we obtain Equation~(\ref{eq:deq}), which is an equation of the fourth order for $\psi_1$.

Note that this complex equation contains two real equations for  the real component $\psi_1$. In~particular, these two equations imply current conservation. The~current conservation can be written in the following form~\cite{Akhmeteli-JMP,AkhmeteliJMParx}:
\begin{equation}\label{eq:d13new}
\textrm{Im}(\psi_4^* \delta)=0,
\end{equation}
where $\delta$ is the left-hand side of Equation~(\ref{eq:deq}).

Obviously, Equation~(\ref{eq:deq}) also implies the following equation:
\begin{equation}\label{eq:d13a}
\textrm{Im}\left(\left(i F^1+F^2\right) \delta\right)=0.
\end{equation}
\hl{One} can check that this equation is a PDE of the third order with respect to $\psi_1$.

In a general case, $\textrm{Im}\left(\frac{i F^1+F^2}{\psi_4^*}\right)$ does not vanish identically, so the system containing Equations~(\ref{eq:d13new}) and (\ref{eq:d13a}) is equivalent to Equation~(\ref{eq:deq}). On~the other hand, current \mbox{conservation (\ref{eq:d13new})} follows also from the Maxwell Equation~(\ref{eq:pr26qr24}). Thus, if~we eliminate all spinor components except $\psi_1$ from Equation~(\ref{eq:pr26qr24}) and add Equation~(\ref{eq:d13a}), we will obtain a system (A) of five real PDEs of the third order with respect to the real indeterminate $\psi_1$, and~this system of equations is generally equivalent to the equations of spinor \mbox{electrodynamics (\ref{eq:pr25qr24}), (\ref{eq:pr25aqr24}), (\ref{eq:pr26qr24})} under the gauge~condition.

It is well-known that, if~there are $d$ independent variables, the~number of all different derivatives of order $r$, where $0\leq r \leq m$, equals the number of combinations
\begin{equation}\label{eq:dcom}
\left( \begin{array}{c} m+d\\ d\end{array}\right)=\frac{(m+d)!}{m!d!}.
\end{equation}
\hl{In} our case, $d=4$ (the number of dimensions of space-time). If~we apply all different derivatives of order $r$, where $0\leq r \leq 4$, to~the equations of system (A), we will obtain a system (B) containing
\begin{equation}\label{eq:dcom2}
5\frac{(4+4)!}{4!4!}=350
\end{equation}
equations. As~the equations of system (A) are of the third order with respect to $\psi_1$, system (B) will contain, in~the worst case, all different derivatives of $\psi_1$ of order $r$, where $0\leq r \leq 7$. Therefore, system (B) will contain no more than
\begin{equation}\label{eq:dcom3}
\frac{(7+4)!}{7!4!}=330
\end{equation}
such derivatives, which we can regard as indeterminates of a system of polynomial equations. Therefore, there are more equations than indeterminates in system (B), so, hopefully, it is possible to use the system to express $\psi_1$ via components of the four-potential of electromagnetic field and their derivatives and thus eliminate the spinor field from spinor electrodynamics. The~above estimates may be naive, and~more rigorous estimates, such as those of~\cite{Gust} may be~required.

\subsection{Lagrangian of Spinor Electrodynamics with Just One Real Function to Describe Charged Spinor~Field}\label{n25}

As three out of four components of the Dirac spinor can be eliminated from the Dirac equation, let us express the Lagrangian of spinor electrodynamics via just one component of the spinor field~\cite{Akhmeteliym}. The~Lagrangian of spinor electrodynamics is~\cite{Bogo,Itzykson}
\begin{equation}\label{eq:lag}
\mathcal{L}=\frac{i}{2}\left(\bar{\psi}\gamma^\mu\psi_{,\mu}-\bar{\psi}_{,\mu}\gamma^\mu\psi\right)-\bar{\psi}\psi-\frac{1}{4 e^2}F_{\mu\nu}F^{\mu\nu}-A_\mu\bar{\psi}\gamma^\mu\psi.
\end{equation}
\hl{Let} us derive the equations of motion for Lagrangian
\vspace{-9pt}
%\begin{adjustwidth}{-\extralength}{0cm}
%\centering %% If there is a figure in wide page, please release command \centering
\begin{eqnarray}\label{eq:lagrprim}
\nonumber
\mathcal{L}'=\mathcal{L}-i\lambda\left(\psi_1^*-\psi_1\right)=\\
\nonumber
\frac{1}{2}\bar{\psi}(i\gamma^\mu\psi_{,\mu}-A_\mu\gamma^\mu\psi-\psi)-
\frac{1}{2}(i\bar{\psi}_{,\mu}\gamma^\mu+A_\mu\bar{\psi}\gamma^\mu+\bar{\psi})\psi-
\\
\frac{1}{4 e^2}F_{\mu\nu}F^{\mu\nu}-i\lambda(\psi_1^*-\psi_1),
\end{eqnarray}
%\end{adjustwidth}
where the Lagrange multiplier $\lambda=\lambda(x)$ is real, under~the condition $\psi_1^*-\psi_1$=0. It is not difficult to obtain:
\begin{equation}\label{eq:pr25d}
(i\slashed{\partial}-\slashed{A}-1)\psi-i\gamma^0\Lambda=0,
\end{equation}
\begin{equation}\label{eq:pr25ad}
\bar{\psi}(i\overleftarrow{\slashed{\partial}}+\slashed{A}+1)-i\Lambda^T=0,
\end{equation}
where
\begin{equation}\label{eq:lambdad2}
\Lambda=\left( \begin{array}{c}
\lambda\\
0\\
0\\
0\end{array}\right),\Lambda^T= (\lambda,0,0,0),
\end{equation}
and the Maxwell Equation~(\ref{eq:pr26qr24}) does not~change.

Let us add Equation~(\ref{eq:pr25d}) multiplied by $\bar{\psi}$ from the left and Equation~(\ref{eq:pr25ad}) multiplied by $\psi$ from the left:
\begin{equation}\label{eq:sum}
i\bar{\psi}\overleftarrow{\slashed{\partial}}\psi+i\bar{\psi}\slashed{\partial}\psi-
i\lambda(\psi_1^*+\psi_1)=0.
\end{equation}
\hl{On} the other hand,
\begin{equation}\label{eq:curpres}
i\bar{\psi}\overleftarrow{\slashed{\partial}}\psi+i\bar{\psi}\slashed{\partial}\psi=
(\bar{\psi}\gamma^\mu\psi)_{,\mu}=0,
\end{equation}
as the Maxwell Equation~(\ref{eq:pr26qr24}) implies current preservation, and, according to the gauge condition, $\psi_1^*=\psi_1$, so $\lambda\psi_1=0$. As~in a general case $\psi_1$ does not vanish identically, we obtain $\lambda=0$. Therefore, Lagrangian $\mathcal{L}'$ (Equation (\ref{eq:lagrprim})) under the condition $\psi_1^*-\psi_1$~=~0 implies the same Dirac equation and Maxwell equation as Lagrangian $\mathcal{L}$ (Equation (\ref{eq:lag})) of spinor electrodynamics, but~in a fixed gauge. Therefore, for~every solution of the equations of spinor electrodynamics, there is a physically equivalent solution for Lagrangian $\mathcal{L}'$ with the gauge condition (a solution that only differs in the choice of gauge). It is also obvious that each solution for Lagrangian $\mathcal{L}'$  under the gauge condition is a solution for Lagrangian $\mathcal{L}$. Thus, Lagrangian $\mathcal{L}$ and Lagrangian $\mathcal{L}'$ under the gauge condition describe the same~physics.

Now let us eliminate all spinor components but $\psi_1$ from Lagrangian $\mathcal{L}'$  under the gauge condition using the Dirac equation and $\lambda=0$. Let us note that the second term in Equation~(\ref{eq:lagrprim}) is a complex conjugation of the first term, so let us focus on the first term $\frac{1}{2}\bar{\psi}(i\gamma^\mu\psi_{,\mu}-A_\mu\gamma^\mu\psi-\psi)$.

Using Equations~(\ref{eq:d5qr22}), (\ref{eq:d6qr22}), (\ref{eq:d72qr22}) and (\ref{eq:d82qr22}), one can see that all components of $i\gamma^\mu\psi_{,\mu}-A_\mu\gamma^\mu\psi-\psi$ vanish except for the second one, which equals the left-hand side of (\ref{eq:deq}). As~the second component of $\bar{\psi}$ is $-\psi_4^*$ and
\begin{eqnarray}\label{eq:d8q}
-\psi_4^*=(A^1-i A^2-\partial_2-i \partial_1)\psi_1^*-(A^0+A^3-i\partial_3+i\partial_0)\psi_2^*
\end{eqnarray}
(cf. (\ref{eq:d6qr22})), we obtain the following Lagrangian of spinor electrodynamics with just one real function to describe charged spinor field:
\vspace{-11pt}
%\begin{adjustwidth}{-\extralength}{0cm}
%\centering %% If there is a figure in wide page, please release command \centering
\begin{eqnarray}\label{eq:d8q2}
\nonumber
\mathcal{L}''=
\\
\nonumber
\textrm{Re}(((A^1-i A^2-\partial_2-i \partial_1-
\\
\nonumber
(A^0+A^3-i\partial_3+i\partial_0)(i F^{1*}-F^{2*})^{-1}(\Box''-i F^{3*}))\psi_1)\times
\\
((\Box'-i F^3)(i F^1+F^2)^{-1}(\Box'+i F^3)-i F^1+F^2)\psi_1)-\frac{1}{4 e^2}F_{\mu\nu}F^{\mu\nu},
\end{eqnarray}
%\end{adjustwidth}
where $\psi_1$ is real and
\begin{eqnarray}\label{eq:d8n2c}
\Box''=\partial^\mu\partial_\mu-2i A^\mu\partial_\mu-i A^\mu_{,\mu}-A^\mu A_\mu+1.
\end{eqnarray}
\hl{Simplification} of the Lagrangian and derivation of its relativistically covariant form along the lines of Section \ref{n23} is left for future~work.

\subsection{Spinor~Electrodynamics}\label{n27}

While there are reasons to believe that differential algebraic elimination of the spinor field is possible (see Section \ref{n24}), so far it has only been achieved at the expense of the introduction of the complex four-potential of an electromagnetic field~\cite{Akhmeteli-EPJC}. We are  again using the equations of spinor electrodynamics
\begin{equation}\label{eq:pr25}
(i\slashed{\partial}-\slashed{A})\psi=\psi,
\end{equation}
\begin{equation}\label{eq:pr26}
\Box A_\mu-A^\nu_{,\nu\mu}=e^2\bar{\psi}\gamma_\mu\psi,
\end{equation}
and the notation at the beginning of Section \ref{n22}.

Let us apply the following ``generalized gauge transform'':
\begin{equation}\label{eq:dd1}
\psi=\exp(i\alpha)\varphi,
\end{equation}
\begin{equation}\label{eq:dd1a}
A_\mu=B_\mu-\alpha_{,\mu},
\end{equation}
where the new four-potential $B_\mu$ is complex (cf.~\cite{Mignani}) but~produces the same electromagnetic fields as $A_\mu$), $\alpha=\alpha(x^\mu)=\beta+i\delta$, $\beta=\beta(x^\mu)$, $\delta=\delta(x^\mu)$, and~$\beta$, $\delta$ are real. The~imaginary part of the complex four-potential is a gradient of a certain function, so alternatively we can use this function instead of the imaginary components of the~four-potential.

After the transform, the~equations of spinor electrodynamics can be rewritten as follows:
\begin{equation}\label{eq:dd2}
(i\slashed{\partial}-\slashed{B})\varphi=\varphi,
\end{equation}
\begin{equation}\label{eq:dd3}
\Box B_\mu-B^\nu_{,\nu\mu}=\exp(-2\delta)e^2\bar{\varphi}\gamma_\mu\varphi.
\end{equation}
\hl{If} $\psi$ and $\varphi$ have components
\vspace{6pt}
\begin{equation}\label{eq:dd4}
\varphi=\left( \begin{array}{c}
\varphi_1\\
\varphi_2\\
\varphi_3\\
\varphi_4\end{array}\right),
\psi=\left( \begin{array}{c}
\psi_1\\
\psi_2\\
\psi_3\\
\psi_4\end{array}\right),
\end{equation}
let us fix the ``gauge transform'' of Equations~(\ref{eq:dd1}) and (\ref{eq:dd1a}) somewhat arbitrarily by the following condition:
\begin{equation}\label{eq:dd5}
\varphi_1=\exp(-i\alpha)\psi_1=1.
\end{equation}
\hl{We} then follow the approach and notation of Section \ref{n22}. The~Dirac Equation~(\ref{eq:dd2}) can be written in components as follows:
\begin{eqnarray}\label{eq:dd6}
(B^0+B^3)\varphi_3+(B^1-i B^2)\varphi_4+i(\varphi_{3,3}-i\varphi_{4,2}+\varphi_{4,1}-\varphi_{3,0})=\varphi_1,
\end{eqnarray}
\begin{eqnarray}\label{eq:dd7}
(B^1+i B^2)\varphi_3+(B^0-B^3)\varphi_4-i(\varphi_{4,3}-i\varphi_{3,2}-\varphi_{3,1}+\varphi_{4,0})=\varphi_2,
\end{eqnarray}
\begin{eqnarray}\label{eq:dd8}
(B^0-B^3)\varphi_1-(B^1-i B^2)\varphi_2-i(\varphi_{1,3}-i\varphi_{2,2}+\varphi_{2,1}+\varphi_{1,0})=\varphi_3,
\end{eqnarray}
\begin{eqnarray}\label{eq:dd9}
-(B^1+i B^2)\varphi_1+(B^0+B^3)\varphi_2+i\varphi_{2,3}+\varphi_{1,2}-i(\varphi_{1,1}+\varphi_{2,0})=\varphi_4.
\end{eqnarray}
\hl{Equations}~(\ref{eq:dd9}) and (\ref{eq:dd8}) can be used to express components $\varphi_3,\varphi_4$ via $\varphi_1,\varphi_2$ and eliminate them from Equations~(\ref{eq:dd6}) and (\ref{eq:dd7}). The~resulting equations for $\varphi_1$ and $\varphi_2$ are as follows:
\begin{eqnarray}\label{eq:dd10}
\nonumber
-\varphi_{1,\mu}^{,\mu}+\varphi_2(-i B^1_{,3}-B^2_{,3}+B^0_{,2}+B^3_{,2}+
i(B^0_{,1}+B^3_{,1}+B^1_{,0})+B^2_{,0})+\\
+\varphi_1(-1+B^{\mu} B_{\mu}-i B^{\mu}_{,\mu}+i B^0_{,3}-B^1_{,2}+B^2_{,1}+i B^3_{,0})-2i B^{\mu}\varphi_{1,\mu}=0,
\end{eqnarray}
\begin{eqnarray}\label{eq:dd11}
\nonumber
-\varphi_{2,\mu}^{,\mu}+i\varphi_1( B^1_{,3}+i B^2_{,3}+i B^0_{,2}-i B^3_{,2}+
B^0_{,1}-B^3_{,1}+B^1_{,0}+i B^2_{,0})+\\
+\varphi_2(-1+B^{\mu} B_{\mu}-i( B^{\mu}_{,\mu}+B^0_{,3}+i B^1_{,2}-i B^2_{,1}+ B^3_{,0}))-2i B^{\mu}\varphi_{2,\mu}=0.
\end{eqnarray}

Equation~(\ref{eq:dd10}) can be used to express $\varphi_2$ via $\varphi_1$:
\begin{eqnarray}\label{eq:dd12}
\varphi_2=-\left(i F^1+F^2\right)^{-1}\left(\Box'+i F^3\right)\varphi_1.
\end{eqnarray}
where the modified d'Alembertian $\Box'$ is defined as follows:
\begin{eqnarray}\label{eq:dd14}
\Box'=\partial^\mu\partial_\mu+2 i B^\mu\partial_\mu+i B^\mu_{,\mu}-B^\mu B_\mu+1.
\end{eqnarray}
\hl{Equation}~(\ref{eq:dd11}) can be rewritten as follows:
\begin{eqnarray}\label{eq:dd15}
-\left(\Box'-i F^3\right)\varphi_2-\left(i F^1-F^2\right)\varphi_1=0.
\end{eqnarray}
\hl{The substitution} of $\varphi_2$ from Equation~(\ref{eq:dd12}) into Equation~(\ref{eq:dd11}) yields an equation of the fourth order for $\varphi_1$:
\begin{eqnarray}\label{eq:dd16}
\left(\left(\Box'-i F^3\right)\left(i F^1+F^2\right)^{-1}\left(\Box'+i F^3\right)-i F^1+F^2\right)\varphi_1=0.
\end{eqnarray}

Application of the gauge condition of Equation~(\ref{eq:dd5}) to \hl{Equations}~(\ref{eq:dd12})--(\ref{eq:dd16}) yields the following equations:
\vspace{6pt}
\begin{eqnarray}\label{eq:dd17}
\Box'\varphi_1=i B^\mu_{,\mu}-B^\mu B_\mu+1,
\end{eqnarray}
\begin{eqnarray}\label{eq:dd18}
\varphi_2=-\left(i F^1+F^2\right)^{-1}\left(i B^\mu_{,\mu}-B^\mu B_\mu+1+i F^3\right),
\end{eqnarray}
\begin{eqnarray}\label{eq:dd19}
\left(\Box'-i F^3\right)\left(i F^1+F^2\right)^{-1}\left(i B^\mu_{,\mu}-B^\mu B_\mu+1+i F^3\right)-i F^1+F^2=0,
\end{eqnarray}
\begin{eqnarray}\label{eq:dd20b}
-\left(\Box'-i F^3\right)\varphi_2-\left(i F^1-F^2\right)=0.
\end{eqnarray}
\hl{Obviously}, Equations~(\ref{eq:dd5}), (\ref{eq:dd8}),~(\ref{eq:dd9}) and  (\ref{eq:dd18}) can be used to eliminate spinor $\varphi$ from the equations of spinor electrodynamics (\ref{eq:dd2}) and (\ref{eq:dd3}). It is then possible to eliminate $\delta$ from the resulting equations. Furthermore, it turns out that the equations describe independent dynamics of the (complex four-potential of) electromagnetic field $B^\mu$. More precisely, if~components $B^\mu$ and their temporal derivatives (derivatives with respect to $x^0$) up to the second order $\dot{B}^\mu$ and $\ddot{B}^\mu$ are known at some point in time in the entire 3D space $x^0$=const, the~equations determine the temporal derivatives of the third order $\dddot{B}^\mu$, so the Cauchy problem can be posed, and~the equations can be integrated (at least locally). This statement is proven in~\cite{Akhmeteli-EPJC}.

Thus, the matter field can be eliminated from equations of scalar electrodynamics and spinor electrodynamics, and~the resulting equations describe the independent evolution of the electromagnetic field (see precise wording above). It should be noted that these mathematical results allow different interpretations. For~example, in~the de Broglie---Bohm interpretation, the electromagnetic field, rather than the wave function, can play the role of the guiding field~\cite{Akhmeteli-IJQI}. Alternatively, one can also use the above results to remove the matter field altogether, as~if it were just a ghost field, and~leave just the electromagnetic field in an interpretation. Another interpretation of real charged fields is described in Section \ref{n40}.

\subsection{Algebraic Elimination of Spinor Components from the Dirac Equation in the Yang--Mills~Field}\label{n28}

Surprisingly, the~procedure of Section \ref{n23} can also be successfully applied in the case of the Dirac equation in the Yang--Mills field~\cite{Akhmeteliym}. Let us start with the equation (~\cite{Schwartz}, p.~493) in the following form (the notation is similar to that of Section \ref{n23}):
\begin{equation}\label{eq:pr25qcd}
(i\slashed{\partial}-\slashed{A})\psi=\psi,
\end{equation}
where $A_\mu=A_\mu^a T^a$, $T^a$ are generators of the SU($n$) group. Components $\psi_{i\alpha}$ of the field $\psi$ have an SU($n$) fundamental representation index (``color'' index) $i$ and a spinor index $\alpha$.

Multiplying both sides of Equation~(\ref{eq:pr25qcd}) by $(i\slashed{\partial}-\slashed{A})$ from the left and using notation
\begin{eqnarray}\label{eq:d8n1}
F_{\mu\nu}=i[i\partial_\mu-A_\mu,i\partial_\nu-A_\nu]=A_{\nu,\mu}-A_{\mu,\nu}+i[A_\mu,A_\nu],
\end{eqnarray}
we obtain (cf.~\cite{Schwartz}, p.173):
\vspace{-11pt}
%\begin{adjustwidth}{-\extralength}{0cm}
%\centering %% If there is a figure in wide page, please release command \centering
\begin{eqnarray}\label{eq:li1qcd}
\nonumber
\psi=(i\gamma^\nu\partial_\nu-A_\nu\gamma^\nu)(i\gamma^\mu\partial_\mu-A_\mu\gamma^\mu)\psi=
(i\partial_\nu-A_\nu)(i\partial_\mu-A_\mu)\gamma^\nu\gamma^\mu\psi=
\\
\nonumber
(i\partial_\nu-A_\nu)(i\partial_\mu-A_\mu)\frac{1}{2}(\{\gamma^\nu,\gamma^\mu\}+
[\gamma^\nu,\gamma^\mu])\psi=
\\
\nonumber
((i\partial_\mu-A_\mu)(i\partial^\mu-A^\mu)+
\\
\nonumber
\frac{1}{4}
(\{i\partial_\nu-A_\nu,i\partial_\mu-A_\mu\}+
[i\partial_\nu-A_\nu,i\partial_\mu-A_\mu])[\gamma^\nu,\gamma^\mu])\psi=
\\
\nonumber
((i\partial_\mu-A_\mu)(i\partial^\mu-A^\mu)+\frac{1}{4}
[i\partial_\nu-A_\nu,i\partial_\mu-A_\mu][\gamma^\nu,\gamma^\mu])\psi=
\\
(-\partial^\mu\partial_\mu-2 i A^\mu\partial_\mu-i A^\mu_{,\mu}+A^\mu A_\mu-\frac{1}{2}F_{\nu\mu}\sigma^{\nu\mu})\psi.
\end{eqnarray}
%\end{adjustwidth}
\hl{We} used the fact that the contraction of a symmetric and an antisymmetric tensor~vanishes.

We obtain:
\begin{equation}\label{eq:li3}
(\Box'+F)\psi=0,
\end{equation}
where the modified d'Alembertian $\Box'$ is defined as follows:
\begin{eqnarray}\label{eq:d8n2}
\Box'=\partial^\mu\partial_\mu+2 i A^\mu\partial_\mu+i A^\mu_{,\mu}-A^\mu A_\mu+1=-(i\partial_\mu-A_\mu)(i\partial^\mu-A^\mu)+1,
\end{eqnarray}
and
\begin{equation}\label{eq:li4}
F=\frac{1}{2}F_{\nu\mu}\sigma^{\nu\mu}.
\end{equation}
\hl{Let} us note that $\Box'$ and $F$ are manifestly relativistically~covariant.

We assume that the set of $\gamma$-matrices satisfies the same hermiticity conditions and the charge-conjugation  matrix $C$ is chosen in the same way as in Section  \ref{n23}.

We will consider ``spinor components'' of the field $\psi$ having the form $\bar{\xi}\psi$, where $\xi$ is a Dirac spinor. Such ``spinor components'' have ``color'' components $\xi^*_\alpha\gamma^0_{\alpha\beta}\psi_{i\beta}$. Let us further assume that $\xi$ is a constant spinor  (so it does not depend on the space-time coordinates $x=(x^0,x^1,x^2,x^3)$, and~$\partial_\mu\xi\equiv 0$) and multiply  both sides of Equation~(\ref{eq:li3}) by $\bar{\xi}$ from the~left:
\begin{equation}\label{eq:li7}
\Box'(\bar{\xi}\psi)+\bar{\xi}F\psi=0.
\end{equation}
\hl{To} derive an equation for only one ``spinor component'' $\bar{\xi}\psi$, we need to express $\bar{\xi}F\psi$ via $\bar{\xi}\psi$. To~simplify this task, we demand that $\xi$ is an eigenvector of $\gamma^5$; in~other words, $\xi$ is either right-handed or left-handed (the derivation can also be performed for such constant spinors multiplied by a function of space-time coordinates). This condition is Lorentz-invariant. Indeed, Dirac spinors $\chi$ transform under a Lorentz transformation as follows:
\begin{equation}\label{eq:li7a}
\chi'=\Lambda \chi,
\end{equation}
where matrix $\Lambda$ is non-singular and commutes with $\gamma^5$ if the Lorentz transformation is proper and anticommutes otherwise~\cite{Bogo}. Therefore, if~$\xi$ is an eigenvector of $\gamma^5$, then $\xi'$ is also an eigenvector of $\gamma^5$, although~not necessarily with the same~eigenvalue.

Eigenvalues of $\gamma^5$ equal either $+1$ or $-1$, so $\gamma^5\xi=\pm\xi$. The~linear subspace of eigenvectors of $\gamma^5$ with the same eigenvalue as $\xi$ is two-dimensional, so we can choose another constant spinor $\eta$ that is an eigenvector of $\gamma^5$  with the same eigenvalue as $\xi$ in such a way that $\xi$ and $\eta$ are linearly independent. This choice is Lorentz-covariant, as~matrix $\Lambda$ in Equation~(\ref{eq:li7a}) is~non-singular.

Obviously, we can derive an equation similar to (\ref{eq:li7}) for $\eta$:
\begin{equation}\label{eq:li7f}
\Box'(\bar{\eta}\psi)+\bar{\eta}F\psi=0.
\end{equation}

We can express $F_{\mu\nu}$ as $F_{\mu\nu}^a T^a$, where $T^a$ are the generators of the SU($n$) group, then $F=F^a T^a$, where $F^a=\frac{1}{2}F_{\nu\mu}^a\sigma^{\nu\mu}$. If~$\gamma^5\xi=\pm\xi$, then $\bar{\xi}=\xi^\dag\gamma^0$ is a left eigenvector of $\gamma^5$ with an eigenvalue $\mp 1$, as~\begin{equation}\label{eq:li7b}
\bar{\xi}\gamma^5=\xi^\dag\gamma^0\gamma^5=-\xi^\dag\gamma^5\gamma^0=-(\gamma^5\xi)^\dag\gamma^0=\mp\bar{\xi}.
\end{equation}
\hl{The} same is true for spinors $\bar{\eta}=\eta^\dag\gamma^0$ (the proof is identical to that in (\ref{eq:li7b})), $\bar{\xi}F^a$, and~$\bar{\eta}F^a$, as~$\gamma^5$ commutes with $\sigma^{\mu\nu}$ ~\cite{Itzykson}.
As the subspace of left eigenvectors of $\gamma^5$ with an eigenvalue $\mp 1$ is two-dimensional and includes spinors $\bar{\xi}F^a$, $\bar{\eta}F^a$, $\bar{\xi}$, and~$\bar{\eta}$, where the two latter spinors are linearly independent (otherwise spinors $\xi$ and $\eta$ would not be linearly independent), there exist such $a^a=a^a(x)$, $b^a=b^a(x)$, $a'^a=a'^a(x)$, $b'^a=b'^a(x)$ that
\begin{equation}\label{eq:lin12qcd}
\bar{\xi}F^a=a^a\bar{\xi}+b^a\bar{\eta},
\end{equation}
\begin{equation}\label{eq:lin13qcd}
\bar{\eta}F^a=a'^a\bar{\xi}+b'^a\bar{\eta}.
\end{equation}

For each spinor $\chi$, the charge conjugated spinor
\begin{equation}\label{eq:lin14}
\chi^c=C\bar{\chi}^T
\end{equation}
can be defined, and~it has the same transformation properties under Lorentz transformations as $\chi$ ~\cite{Schweber}. We have
\begin{equation}\label{eq:lin15}
\bar{\chi}\chi^c=\bar{\chi}C\bar{\chi}^T=(\bar{\chi})_\alpha C_{\alpha\beta}(\bar{\chi})_\beta=0,
\end{equation}
as $(\bar{\chi})_\alpha(\bar{\chi})_\beta$ and $ C_{\alpha\beta}$ are, respectively, symmetric and antisymmetric (see Equation~(\ref{eq:li6qr27})) with respect to the transposition of $\alpha$ and $\beta$.

Let us multiply Equations~(\ref{eq:lin12qcd}) and (\ref{eq:lin13qcd}) by $\xi^c$ and $\eta^c$ from the right:
\begin{eqnarray}\label{eq:lin16qcd}
\nonumber
\bar{\xi}F^a\xi^c=a^a(\bar{\xi}\xi^c)+b^a(\bar{\eta}\xi^c)=b^a(\bar{\eta}\xi^c),
\\
\nonumber
\bar{\xi}F^a\eta^c=a^a(\bar{\xi}\eta^c)+b^a(\bar{\eta}\eta^c)=a^a(\bar{\xi}\eta^c),
\\
\nonumber
\bar{\eta}F^a\xi^c=a'^a(\bar{\xi}\xi^c)+b'^a(\bar{\eta}\xi^c)=b'^a(\bar{\eta}\xi^c),
\\
\nonumber
\bar{\eta}F^a\eta^c=a'^a(\bar{\xi}\eta^c)+b'^a(\bar{\eta}\eta^c)=a'^a(\bar{\xi}\eta^c),
\end{eqnarray}
so
\begin{eqnarray}\label{eq:lin17qcd}
a^a=\frac{\bar{\xi}F^a\eta^c}{\bar{\xi}\eta^c},
b^a=\frac{\bar{\xi}F^a\xi^c}{\bar{\eta}\xi^c},
a'^a=\frac{\bar{\eta}F^a\eta^c}{\bar{\xi}\eta^c},
b'^a=\frac{\bar{\eta}F^a\xi^c}{\bar{\eta}\xi^c}.
\end{eqnarray}
\hl{Let} us note that
\begin{equation}\label{eq:lin18}
\bar{\xi}\eta^c=\bar{\xi}C\bar{\eta}^T=(\bar{\xi})_\alpha C_{\alpha\beta}(\bar{\eta})_\beta=-(\bar{\eta})_\beta C_{\beta\alpha}(\bar{\xi})_\alpha=-\bar{\eta}\xi^c
\end{equation}
and
\begin{equation}\label{eq:lin19qcd}
\bar{\xi}F^a\eta^c=\bar{\xi}F^a C\bar{\eta}^T=(\bar{\xi}F^a C\bar{\eta}^T)^T=\bar{\eta}C^T F^{aT}\bar{\xi}^T=\bar{\eta}F^a C\bar{\xi}^T=\bar{\eta}F^a\xi^c,
\end{equation}
as
\begin{equation}\label{eq:lin20}
\sigma_{\mu\nu}C=-C\sigma_{\mu\nu}^T
\end{equation}
(see Equations~(\ref{eq:li5qr27}) and (\ref{eq:li6qr27})). Therefore,
\begin{equation}\label{eq:lin20aqcd}
b'^a=-a^a.
\end{equation}

Equations~(\ref{eq:li7}), (\ref{eq:li7f}), (\ref{eq:lin12qcd}) and (\ref{eq:lin13qcd}) yield
\begin{eqnarray}\label{eq:lin21qcd}
\nonumber
\Box'(\bar{\xi}\psi)+a(\bar{\xi}\psi)+b(\bar{\eta}\psi)=0,
\\
\nonumber
\Box'(\bar{\eta}\psi)+a'(\bar{\xi}\psi)+b'(\bar{\eta}\psi)=\Box'(\bar{\eta}\psi)+a'(\bar{\xi}\psi)-a(\bar{\eta}\psi)=0,
\end{eqnarray}
where
\begin{equation}\label{eq:lin22plqcd}
a=a^a T^a,b=b^a T^a,a'=a^a T^a,b'=b'^a T^a,
\end{equation} so
\begin{equation}\label{eq:lin22qcd}
\bar{\eta}\psi=-b^{-1}(\Box'(\bar{\xi}\psi)+a(\bar{\xi}\psi))
\end{equation}
and
\begin{equation}\label{eq:lin23qcd}
(\Box'-a)(-b^{-1})(\Box'+a)(\bar{\xi}\psi)+a'(\bar{\xi}\psi)=0
\end{equation}
or
\begin{equation}\label{eq:lin24qcd}
((\Box'-a)b^{-1}(\Box'+a)-a')(\bar{\xi}\psi)=0.
\end{equation}
\hl{Substituting} the expressions for $a^a$, $b^a$, $a'^a$ from Equation~(\ref{eq:lin17qcd}) and using Equations~(\ref{eq:lin18}) and (\ref{eq:lin24qcd}), we finally obtain:
\begin{equation}\label{eq:lin25qcd}
(((\bar{\xi}\eta^c)\Box'-\bar{\xi}F\eta^c)(\bar{\xi}F\xi^c)^{-1}((\bar{\xi}\eta^c)\Box'
+\bar{\xi}F\eta^c)+\bar{\eta}F\eta^c)(\bar{\xi}\psi)=0.
\end{equation}

This equation for one ``spinor component'' $\bar{\xi}\psi$ is generally equivalent to the Dirac equation in the Yang--Mills field (\ref{eq:pr25qcd}) (if $\bar{\xi}F\xi^c$ is not identically singular, which is generally the case): on the one hand, it was derived from the Dirac equation;  on~the other hand, the~field $\psi$ can be restored if its ``spinor component'' $\bar{\xi}\psi$ is known, and the Dirac Equation~(\ref{eq:pr25qcd}) can be derived from Equation~(\ref{eq:pr25qcd}). This can be demonstrated along the same lines as in~\cite{Akhm2015,Akhmspr}.

The only ``spinor component'' in Equation~(\ref{eq:lin25qcd}) has $n$ ``color'' components; however, using a gauge transformation, all of them but one can be made zero and the remaining component  can be made real, as~the action of the SU($n$) group on the unit sphere in $\mathbb{C}^n$ is transitive for $n\geq 2$ ~\cite{Knapp}.

\subsection{Plasma-like Model of Quantum~Particles}\label{n40}

There is a well-known analogy between quantum particles and plasma: the dispersion relation for the Klein--Gordon equation ($c=\hbar=1$)
\begin{equation}\label{eq:kgdr}
\omega^2=m^2+k^2
\end{equation}
is similar to the dispersion relation for waves in a simple plasma model
\begin{equation}\label{eq:pdr}
\omega^2=\omega_p^2+k^2.
\end{equation}
\hl{However}, to~expand this analogy, we need a description of both negatively and positively charged particles. The description is based on the possibility of using a real, rather than complex, wave function to describe charged particles. This possibility is described in Sections \ref{n22}--\ref{n24}.

Using one real wave function instead of complex functions introduces some ``symmetry'' between positive and negative frequencies and, thus, particles and antiparticles. Therefore, a~tentative description of such (one-particle electron) wave function  was given in~\cite{Akhm10,Akhmeteli-EPJC,Akhm-Ent}: the wave function can describe $N+1$ electrons and $N$ positrons, where $N$ is very large. This collection of particles and antiparticles has the same total charge (and mass) as an electron, and~the value of the wave function at some point (or some combination of the wave function and its derivatives at the point) is a measure of both  ``vacuum polarization'' and the density of probability of finding an electron at this point (finding a positron at that point is also possible, but~probably requires much more energy). An~electron found during a measurement can be any of the $N+1$ electrons. The~results of the measurement on the specific collection can depend, say, on~the precise coordinates of the particles in the collection. One can consider such a collection as a ``dressed'' electron with a well-defined total charge. The~description assumes trajectories of the ``bare'' electrons and positrons, but~the uncertainty principle is valid for the ``dressed'' electron. The~charge density distribution of the ``dressed'' electron is defined by all charges of the ``bare'' electrons and positrons and can be very close to the charge density distribution built from the traditional wave function (see below in this subsection). If~an electron is removed from the collection (for example, a~precise position measurement means high momentum uncertainty, and as~a result, some particle acquires high speed and quickly leaves the collection; the area around the place vacated by the particle will tend to be filled with the surrounding particles) and the energy of the remaining collection is not high enough for the collection to manifest as pairs, it is difficult to tell the collection from the vacuum. This may be the source of discreteness emphasized in~\cite{Khentropy}. It is also important to note that the spreading of wave packets (which complicates, e.g.,~the de Broglie's double solution approach~\cite{colin7}) is not problematic for this~description.

The description of this article only covers the unitary evolution of quantum theory but~not the wave function collapse. The~rationale for that is provided in the introduction of~\cite{Akhm-Ent}.

One can object that the mass of such a collection of a large number of particles and antiparticles would be too large, as~each pair should have a mass of at least two electron masses, but~it is not necessarily so, as~the energy of an electron and a positron that are very close to each other can be significantly less than their energy when the distance between them is~large.

\begin{figure}[H]
 \begin{tabular}{cc}
\includegraphics[width=0.45\linewidth]{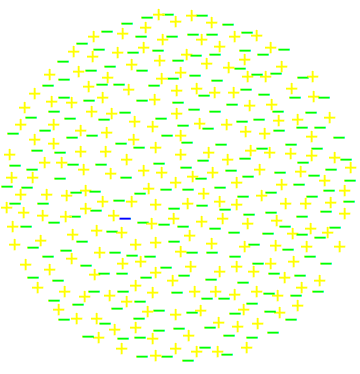}
&\includegraphics[width=0.45\linewidth]{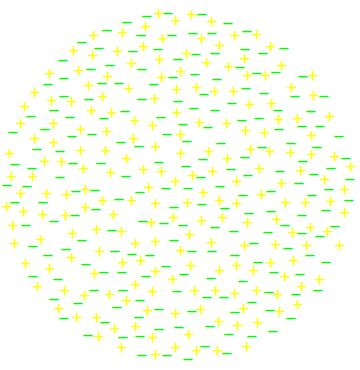}\\
({\bf a})&({\bf b})\\
\end{tabular}
\caption{Collections (\textbf{a},\textbf{b}) have 201 and 200 electrons, respectively, and~200 positrons~each.}
\label{fig:pm}
\end{figure}

The above description is illustrated in Figure~\ref{fig:pm}, where electrons and positrons are represented by minus and plus signs, respectively. Collections (a) and (b) there are identical except for an extra electron (represented by a blue minus sign) in collection (a). The~collections are difficult to visually distinguish from each other, but~the total charges of the collections are one and zero electron charge, respectively. Figure~\ref{fig:pm} is similar to M. Strassler's Figure~3 in~\cite{Strassler}, but~the figure here describes an electron, rather than a nucleon (which is a composite particle), and~the size of the collection is defined by the size of the volume where the wave function does not~vanish.

Similarities of the plasma-like description with the Bohmian interpretation are described in~\cite{Akhm-Ent}.

While single-particle systems are very important (for example, they are sufficient to describe the double-slit experiment), it is necessary to discuss many-particle systems. While the~author does not have a complete description of second quantization for such systems, the~Pauli principle for fermions can emerge in the following way: for identical wave functions, the~relevant collections of discrete charges have identical or very similar coordinates of the charges, and thus a combination of such collections can have very high energy, for~example, due to Coulomb interaction. Matter and radiation (and, eventually, Fermi and Bose fields) are treated differently in the plasma-like description. The second quantization for bosons can be performed using the approach described in Section \ref{n34}.

%\unskip

How accurately can a continuous charge density distribution for a specific wave function with a total charge equal to one electron charge be approximated by a collection of discrete charges with values of $\pm 1$ electron charge? It is obvious that a Fourier expansion of a point-like charge density distribution contains arbitrarily high spatial frequencies, whereas high-spatial-frequency Fourier components of smooth charge density distributions tend to zero fast; therefore, it is probably impossible to approximate a continuous charge density distribution by a finite number of discrete quantized charges with good accuracy. However, quantum field theories are typically considered to be just effective theories. Therefore, we can look for collections of discrete charges with values of $\pm 1$ electron charge that have the same Fourier components with spatial frequencies below some limit value as the initial smooth charge density~distribution.

Let us illustrate this approach in the one-dimensional case assuming that the smooth charge density distribution is periodic, e.g.,~with a period of $2\pi$, so we consider it on a segment $[-\pi,\pi]$. Let us choose some normalized function $f(x)$:
\begin{equation}\label{eq:norm}
\int_{-\pi}^\pi f(x)dx=1
\end{equation}
(the total charge in the charge density distribution is +1 (electron charge)).
The charge density distribution does not have to be non-negative everywhere, as~we have in mind applications not just to the Schr{\"o}dinger equation and the Dirac equation but~also to the Klein--Gordon equation, where the charge density does not have to be of the same sign~everywhere.

Let us consider the Fourier expansion:
\begin{equation}\label{eq:four}
f(x)=\frac{1}{2}a_0+\sum_{k=1}^\infty(a_k \cos (k x)+b_k \sin (k x)),
\end{equation}
where
\begin{equation}\label{eq:fourser}
a_k=\frac{1}{\pi}\int_{-\pi}^\pi f(y)\cos (k y) dy,\;\;b_k=\frac{1}{\pi}\int_{-\pi}^\pi f(y)\sin (k y) dy.
\end{equation}

We will try to find a discrete charge density distribution approximating $f(x)$. Let us assume that this distribution describes $2j+1$ particles with coordinates $x_n$, including $j+1$ electrons and $j$ positrons, so the discrete charge density distribution is
\begin{equation}\label{eq:gx}
g(x)=\sum_{n=1}^{2j+1}(-1)^{n+1}\delta(x-x_n).
\end{equation}

Let us demand that the $k$-th cosine Fourier components of distribution $g(x)$ coincide with $a_k$ for $1\leq k\leq k_c$ (the zeroth cosine Fourier component of $g(x)$ automatically coincides with $a_0$ due to Equation~(\ref{eq:norm})) and that the $k$-th sine Fourier components of distribution $g(x)$ coincide with $b_k$ for $1\leq k\leq k_s$, where $k_c$ and $k_s$ are some natural~numbers.

To find such distribution, one needs to solve a system of polynomial equations~\cite{Akhm-Ent}. It is possible to do that numerically, using the powerful homotopy continuation method for polynomial systems (see~\cite{vers} and references there). That was achieved in~\cite{Akhm-Ent} for a system of 98~equations.

One can look for discrete distributions containing a number of points that is much larger than the number of Fourier coefficients that need to be equal for the smooth and the discrete distributions, so it was conjectured~\cite{Akhm-Ent} (but not proven) that there is always a real solution, no matter how many Fourier coefficients need to be equal. We prove this here for the one-dimensional~case.

Let $f(t)$, where $-\pi\leq t<\pi$, be a real function with a Fourier series
\begin{equation}\label{eq:n21}
f(t)=\sum_{k=-\infty}^{\infty}c_k e^{ikt}, c_k=c_{-k}^*=\frac{1}{2\pi}\int_{-\pi}^{\pi} f(\tau)e^{-i k \tau}d\tau,
\end{equation}
and
\begin{equation}\label{eq:n21a}
\int_{-\pi}^{\pi} f(\tau)d\tau=1,
\end{equation}
so $c_0=\frac{1}{2\pi}$.

Then let us prove the following statement (A): $\forall l\geq 0$ $\exists N,$ $t_j$  $(N\geq 0,$ $-N\leq j\leq N,$ $-\pi< t_j<\pi)$, such that the distribution
\begin{equation}\label{eq:n22}
D_l(t)=\delta(t-t_0)+\sum_{j=-N}^{N}\sgn(j)\delta(t-t_j)
\end{equation}
has the same partial Fourier sum
\begin{equation}\label{eq:n23}
\sum_{k=-l}^{l}c_k e^{ikt}
\end{equation}
as the function $f(t)$.

Let us define a crystal of order $m$ as a distribution
\begin{equation}\label{eq:n24}
C_m(t)=\sum_{n=0}^{m-1}\left(\delta\left(t+\pi-n\frac{2\pi}{m}-\alpha\right)-\delta\left(t+\pi-n\frac{2\pi}{m}-\beta\right)\right),
\end{equation}
where $0<\alpha, \beta<\frac{2\pi}{m}$.
Intuitively, a~``crystal'' contains an equidistant ``comb'' of $m$ positively charged point particles and an equidistant ``comb'' of $m$ negatively charged particles. The~``positive'' comb is displaced with respect to the ``negative''~comb.

Let us prove the above statement (A) by building the required distribution $D_l(t)$ inductively. We choose $D_0(t)$ as $\delta(t-t_0)$, where $t_0$ is an arbitrary number such that $-\pi< t_0<\pi$. Let us assume that we have a distribution $D_{l-1}(t)$ ($l>0$) that is a sum of $\delta(t-t_0)$ and a non-negative integer number of crystals whose orders are less than $l$, and~$D_{l-1}(t)$ has the same partial Fourier sum
\begin{equation}\label{eq:n25}
\sum_{k=-l+1}^{l-1}c_k e^{ikt}
\end{equation}
as the function $f(t)$ (distribution $D_0(t)$ satisfies these conditions for $l=1$).

To build the distribution $D_l(t)$, let us calculate the Fourier coefficients $c_{k m}$ of $C_m(t)$, the~crystal of order $m$ (eq. (\ref{eq:n24})):
%\keywords{quantum, plasma, antiparticle, homotopy continuation} % Write down at least 3 Keywords
\begin{eqnarray}\label{eq:n26}
c_{k\, m}=
\\
\nonumber
\frac{1}{2\pi}\int_{-\pi}^{\pi} \sum_{n=0}^{m-1}\left(\delta\left(\tau+\pi-n\frac{2\pi}{m}-\alpha\right)-
\delta\left(\tau+\pi-n\frac{2\pi}{m}-\beta\right)\right)e^{-i k \tau}d\tau=
\\
\nonumber
\frac{1}{2\pi}\sum_{n=0}^{m-1}\left(e^{-i k \left(-\pi+n\frac{2\pi}{m}+\alpha\right)}-e^{-i k \left(-\pi+n\frac{2\pi}{m}+\beta\right)}\right)=
\\
\nonumber
\frac{1}{2\pi}\left(e^{-i k \left(-\pi+\alpha\right)}-e^{-i k \left(-\pi+\beta\right)}\right)\sum_{n=0}^{m-1}e^{-i kn\frac{2\pi}{m}}.
\end{eqnarray}
\hl{It} is obvious that $c_{0\,m}=0$. We also \hl{obtain}:
\begin{eqnarray}\label{eq:n27}
e^{-i k \left(-\pi+\alpha\right)}-e^{-i k \left(-\pi+\beta\right)}=e^{i k\pi}\left(e^{-ik\alpha}-e^{-ik\beta}\right)=
\\
\nonumber
e^{i k\pi}e^{-ik\frac{\alpha+\beta}{2}}\left(e^{-ik\frac{\alpha-\beta}{2}}-e^{-ik\frac{\beta-\alpha}{2}}\right)=
\\
\nonumber
=2 i e^{i k\pi}e^{-ik\frac{\alpha+\beta}{2}}\sin \left(k\frac{\beta-\alpha}{2}\right).
\end{eqnarray}
\hl{As} the sum
\begin{equation}\label{eq:n28}
\sum_{n=0}^{m-1}e^{-i kn\frac{2\pi}{m}}
\end{equation}
does not change when multiplied by $e^{-i k\frac{2\pi}{m}}$, it vanishes if $1\leq |k|\leq m-1$ and equals $m$ if $k=0$ or $|k|=m$ (we do not need the explicit expressions for the sum for $|k|>m$). Thus, $c_{k\, m}$ vanishes if $|k| \leq m-1$, and~\begin{equation}\label{eq:n29}
c_{k\, m}=\frac{i m}{\pi} e^{i k\pi}e^{-ik\frac{\alpha+\beta}{2}}\sin \left(k\frac{\beta-\alpha}{2}\right)
\end{equation}
if $|k|=m$.

Therefore, if~we add an integer number of crystals $C_m(t)$ to the distribution $D_{l-1}(t)$, the~Fourier components of the latter will not change for $|k|<m$ and will coincide with those for $f(t)$. Let us prove that $\forall$ complex $z$ $\exists$ $\alpha,\beta$  $(0<\alpha, \beta<\frac{2\pi}{m})$, and~integer $j$ such that
\vspace{6pt}
\begin{equation}\label{eq:n210}
j c_{m\, m}=j \frac{i m}{\pi} e^{i m\pi}e^{-im\frac{\alpha+\beta}{2}}\sin \left(m\frac{\beta-\alpha}{2}\right)=z.
\end{equation}
To do this, it is sufficient to prove that $\forall$ complex $z_1=\rho e^{i\varphi}$ $(0\leq\rho\leq\frac{1}{2}, -\frac{3}{2}\pi\leq \phi <-\frac{\pi}{2})$ $\exists$ $\alpha,\beta$  $(0\leq\alpha, \beta<\frac{2\pi}{m})$, such that
\begin{equation}\label{eq:n211}
e^{-im\frac{\alpha+\beta}{2}}\sin \left(m\frac{\beta-\alpha}{2}\right)=z_1.
\end{equation}
\hl{After} that, we will be able to choose $z_1$ and the sign and the magnitude of $j$ in such a way~that
\begin{equation}\label{eq:n211a}
z=j \frac{i m}{\pi} e^{i m\pi}z_1,
\end{equation}
and Equation~(\ref{eq:n210}) is~satisfied.

We can find the required values of $\alpha$ and $\beta$ from the following equations:
\begin{eqnarray}\label{eq:n212}
-m\frac{\alpha+\beta}{2}=\varphi,
\nonumber
\\
m\frac{\beta-\alpha}{2}=\varepsilon,
\end{eqnarray}
where $\varepsilon$ is such that $\sin\varepsilon=\rho$, $0\leq\epsilon\leq\frac{\pi}{6}$, and~$-\frac{3}{2}\pi\leq \phi <-\frac{\pi}{2}$. We obtain: $\alpha=\frac{-\phi-\varepsilon}{m}$, $\beta=\frac{-\phi+\varepsilon}{m}$, and~$\frac{\pi}{3 m}<\alpha\leq\frac{3\pi}{2 m}$, $\frac{\pi}{2 m}<\beta\leq\frac{5\pi}{3 m}$, so $0<\alpha, \beta<\frac{2\pi}{m}$, and~Equation~(\ref{eq:n210}) is~proven.

If $D_{l-1}(t)$ has the following Fourier series:
\begin{equation}\label{eq:n213}
D_{l-1}(t)=\sum_{k=-\infty}^{\infty}d_{k\, l-1} e^{ikt}
\end{equation}
(where $d_{k\, l-1}=c_k$ for $-l+1\leq k\leq l-1$), let us choose an integer $j$ and a crystal $C_l(t)$ in such a way that $j c_{l\, l}=c_l-d_{l\, l-1}$. Let us then consider the following distribution:
\begin{equation}\label{eq:n214}
D_l(t)=D_{l-1}(t)+j C_l(t).
\end{equation}
\hl{Its} Fourier components $d_{l\, k}$ coincide with those of $f(t)$ for $-l+1\leq k\leq l-1$ and for $k=l$. As~$f(t)$, $D_{l-1}(t)$, and~$D_l(t)$ are real and, therefore, $c_k=c_{-k}^*$, $d_{k\, l-1}=d_{-k\, l-1}^*$ and $d_{k\, l}=d_{-k\, l}^*$, we also have $c_{-l}=d_{-l\, l}$. Therefore, statement (A) is~proven.

It is not clear if a similar approach can be used to find a discrete distribution with all Fourier components coinciding with those of the smooth distribution (to this end, the~summation over $j$ in the right-hand side of Equation~(\ref{eq:n22}) would be from $-\infty$ to $+\infty$).

The approach of this section can be useful for the interpretation of quantum phase-space distribution functions~\cite{lee}, such as the Wigner distribution function, which is not necessarily non-negative. It is also interesting to compare the approach with the initial Schrödinger's interpretation of the wave function (see, e.g.,~~in~\cite{Sebens,Barutschr} and references therein). Schrödinger's interpretation of $e|\psi|^2$ as charge density meets some objections. For~example, A. Khrennikov noted (\cite{Khbq}, p.~23) :``Unfortunately, I was not able to find in Schrödinger's papers any explanation of the impossibility to divide this cloud into a few smaller clouds, i.e.,~no attempt to explain the fundamental discreteness of the electric charge.'' The plasma-like description suggests that $e|\psi|^2$ (and its analogs for the Klein--Gordon and Dirac equations) is just a smoothed charge density, and~the description is immune to the above~objection. Note that the negativity of the Wigner function is considered a resource for quantum computing~\cite{Booth}.

An example mathematical model (equations of motion) of collections of charged particles and antiparticles interacting with an electromagnetic field was built in~\cite{Akhm-Ent}. The model is based on the results described in Section \ref{n21} and emulates arbitrarily well (for a sufficiently large cutoff constant) a quantum theory---the Klein--Gordon--Maxwell electrodynamics (scalar electrodynamics); therefore, one can reasonably expect that the model should successfully describe a wide spectrum of quantum phenomena. Modeling of specific experiments, such as the double-slit experiment, is left for future~work.

Let us emphasize that one needs to choose the cutoff constant to fully define the mathematical model. While this constant can be arbitrarily high, the~model has problems at very high temporal/spatial frequencies once this choice is made. However, these problems seem similar to those of standard quantum field theories~\cite{georgi}.

The following problem of the plasma-like description was resolved in~\cite{Akhm-Ent}. Composite particles, such as nucleons or large molecules, also demonstrate quantum properties~\cite{zeilneut,Fein}. It is, however, difficult to imagine that molecule--antimolecule pairs play a significant role in diffraction of large molecules (creation of such pairs is possible, but~much less probable than the creation of electron--positron pairs). However, composite particles take part in some interactions (for example, electromagnetic or strong interactions), so the description can be modified as follows in that case: composite particles are accompanied by a large collection of fermion--antifermion pairs (for example, electron--positron pairs for electromagnetic interactions and quark--antiquark pairs for strong interactions; in some situations, it can be difficult to tell such pairs from force carriers, such as photons or gluons). Such fermions/pairs/force carriers are present at all locations where the wave function traditionally describing the composite particle does not vanish, so the dimensions of the collection are not limited by the range of the interaction (for example, the~short range of strong interaction). Thus, the~composite particle can be detected at all locations where the wave function does not vanish, although~at most locations, it is fermions/pairs/force carriers of the collection that interact directly with the instrument, not the composite particle~itself.

There is an obvious analogy between this description and plasma. As~the dispersion relation for the Klein--Gordon equation $\omega^2=m^2+k^2$ (in a natural system of units) is similar to a dispersion relation of a simple plasma model, such analogy was used previously (see, e.g.,~in~\cite{Vig,Plyu,Shi}). This analogy illustrates the effective long-range interaction within a~collection.

In~\cite{Akhm-Ent}, a~preliminary estimate was made  of the density of particles in a collection modeling, which is perceived as one particle in traditional quantum experiments. If~$n_e$ is the electron density in the collection, the~plasma frequency $\omega_p$ in the electron--positron plasma is $\sqrt{2}$ times greater than the traditional plasma frequency~\cite{stenson}, i.e.,
\begin{equation}\label{eq:pf}
\omega_p=\sqrt{\frac{8\pi n_e e^2}{m_e}}
\end{equation}
(we do not consider any renormalization of mass and charge in this preliminary treatment). It is natural to suggest that this plasma frequency is equal (maybe on the order of magnitude) to the angular frequency of Zitterbewegung $\frac{2m_e c^2}{\hbar}$ ~\cite{Diraczit,thallerzit}, so we obtain
\begin{equation}\label{eq:pf2}
n_e=\frac{m_e^3 c^4}{2\pi\hbar^2 e^2}=\left(\frac{m_e c}{\hbar}\right)^3\frac{c\hbar}{2\pi e^2}=\frac{1}{2\pi\alpha}\left(\frac{\hbar}{m_e c}\right)^{-3},
\end{equation}
where $\alpha=\frac{e^2}{\hbar c}$ is the fine structure constant. Thus, $n_e\approx 3.8\cdot10^{32}$ cm$^{-3}$ or 21.8 per cube with an edge length equal to the reduced Compton wavelength $\frac{\hbar}{m_e c}\approx 3.86\cdot 10^{-11}~\textrm{cm}$. The~high electron density suggests that there is low energy per particle of a collection. Let us also note that in this context, the Zitterbewegung frequency plays a role of a ``natural frequency'', rather than a frequency of some ``internal clock''~\cite{Gou}.

\subsection{Transition to Many-Particle~Theories}\label{n34}

To make this article more self-contained, this section contains a summary of the approach used in~\cite{Akhmeteli-IJQI} (following~\cite{nightnew}) to embed the resulting (non-second-quantized) theories describing the independent evolution of an electromagnetic field in quantum field theories. The~following off-the-shelf mathematical result~\cite{Kowalski,Kowalski2}, a~generalization of the Carleman linearization, generates for a system of nonlinear partial differential equations a system of linear equations in the Fock space, which looks like a second-quantized theory and is equivalent to the original nonlinear system on the set of solutions of the~latter.

Let us consider a nonlinear differential equation in an (s + 1)-dimensional space-time (the equations describing independent dynamics of electromagnetic field for scalar electrodynamics and spinor electrodynamics are a special case of this \hl{equation})\linebreak ${\partial_t}\boldsymbol{\xi}(x,t) = \boldsymbol{F}(\boldsymbol{\xi},{D^\alpha}\boldsymbol{\xi};x,t)$, $\boldsymbol{\xi}(x,0)=\boldsymbol{\xi}_0(x)$, where $\boldsymbol{\xi}:\mathbf{R}^s\times\mathbf{R}\rightarrow\mathbf{C}^k$ (function $\boldsymbol{\xi}$ is defined in an (s+1)-dimensional space-time and takes values in a $k$-dimensional complex space; for example, for~spinor electrodynamics, the~space-time is (3+1)-dimensional, and~$\boldsymbol{\xi}$ includes real and imaginary parts of $B^\mu$, $\dot{B}^\mu$, and~$\ddot{B}^\mu$),  $D^\alpha\boldsymbol{\xi}=\left(D^{\alpha_1}\xi_1,\ldots ,D^{\alpha_k}\xi_k\right)$, $\alpha_i$ are multiindices, $ {D^\beta}={\partial^{|\beta|}}/\partial x_1^{\beta_1}\ldots\partial x_s^{\beta_s}$, with~$ |\beta|=\sum\limits_{i=1}^{s}\beta_i$, is a generalized derivative, $\boldsymbol{F}$ is analytic in $\boldsymbol{\xi}$, $D^\alpha\boldsymbol{\xi}$. It is also assumed that $\boldsymbol{\xi_0}$ and $\boldsymbol{\xi}$ are square integrable. Then Bose operators $\boldsymbol{a^\dag(x)}=\left(a^\dag_1(x),\ldots,a^\dag_k(x)\right)$ and $\boldsymbol{a(x)}=\left(a_1(x),\ldots,a_k(x)\right)$ are introduced with the canonical commutation relations:
\begin{eqnarray}\label{eq:ad1}
\nonumber
\left[a_i(x),a^\dag_j(x')\right]=\delta_{ij}\delta(x-x')I,\\
\left[a_i(x),a_j(x')\right]=\left[a^\dag_i(x),a^\dag_j(x')\right]=0,
\end{eqnarray}
where $x,x'\in\mathbf{R}^s$, $i,j=1,\ldots,k$. Normalized functional coherent states in the Fock space are defined as $|\boldsymbol{\xi}\rangle =\exp\left(-\frac{1}{2}\int d^sx|\boldsymbol{\xi}|^2\right)\exp\left(\int d^sx\boldsymbol{\xi}(x)\cdot\boldsymbol{a}^\dagger(x)\right)|\boldsymbol{0}\rangle$. They have the following property:
\begin{equation}\label{eq:ad1a}
\boldsymbol{a}(x)|\boldsymbol{\xi}\rangle =\boldsymbol{\xi}(x)|\boldsymbol{\xi}\rangle.
\end{equation}
\hl{Then} the following vectors in the Fock space can be  introduced:
\begin{eqnarray}\label{eq:ad2}
\nonumber
|\xi,t\rangle = \exp\left[\frac{1}{2}\left(\int {d^s}x|\boldsymbol{\xi}|^2-\int {d^s}x|\boldsymbol{\xi}_0|^2\right)\right]|\boldsymbol{\xi}\rangle=\\
\exp\left(-\frac{1}{2}\int d^sx|\boldsymbol{\xi}_0|^2\right)
\exp\left(\int d^sx\boldsymbol{\xi}(x)\cdot\boldsymbol{a}^\dagger(x)\right)|\boldsymbol{0}\rangle.
\end{eqnarray}
\hl{Differentiation} of Equation~(\ref{eq:ad2}) with respect to time $t$ yields, together with Equation~(\ref{eq:ad1a}), a~linear Schr\"{o}dinger-like evolution equation in the Fock space:
\begin{eqnarray}\label{eq:ad3}
\frac{d}{dt}|\xi,t\rangle = M(t)|\xi,t\rangle,
|\xi,0\rangle=|\boldsymbol{\xi}_0\rangle,
\end{eqnarray}
where the boson ``Hamiltonian''
$M(t) = \int {d^s}x{\boldsymbol{a}^\dagger}(x)\cdot F(\boldsymbol{a}(x),{D^\alpha}\boldsymbol{a}(x))$. Let us note that the states of Equation~(\ref{eq:ad2}) are, in general, multi-particle~states.

Applicability of the principle of superposition in this approach is discussed in~\cite{Akhmeteli-EPJC}.

\vspace{6pt}

\section{Conclusions}

We considered some mathematical results that enable classical models of quantum gauge theories and thus seem relevant to the interpretation of quantum mechanics. Some of the results, such as the algebraic elimination of three out of four spinor components from the Dirac equation in an  electromagnetic field and the proof of independent evolution of an electromagnetic field in scalar electrodynamics, are valuable regardless of~interpretation.

\vspace{6pt}
%%%%%%%%%%%%%%%%%%%%%%%%%%%%%%%%%%%%%%%%%%

\section*{Funding}
This research received no external~funding.

\section*{Acknowledgements}
The author is grateful to A. V. Gavrilin, A. Yu. Kamenshchik, T. G. Khunjua, A. D. Shatkus, and~ A. Tarasevitch for their interest in this work and valuable~remarks.

\bibliographystyle{unsrt}

\end{document}